# A quantitative aspect of the Heisenberg uncertainty.

A. K. Khitrin

*Department of Chemistry, Kent State University*

**Abstract.**
Minimization of the expectation value of energy under the constraints imposed by the uncertainty principle can be a convenient method of solving quantum-mechanical problems.

1. **Introduction**

The uncertainty principle, formulated by Heisenberg [1] as

$$\Delta r_x \, \Delta p_x \sim h, \qquad (1)$$

where $\Delta r_x$ and $\Delta p_x$ are uncertainties in coordinate and momentum, and $h$ is the Plank's constant, is a consequence of wave nature of matter and the superposition principle. It is often used to illustrate qualitatively the differences between quantum and classical mechanics. More accurate lower bound for the product of uncertainties is (Kennard [2], Weyl [3])

$$<r_x^2> <p_x^2> \geq (\hbar/2)^2. \qquad (2)$$

For arbitrary quantum-mechanical operators $A$ and $B$ the uncertainty relation is (Robertson [4], Schrödinger [5])

$$<A^2> <B^2> \geq <i[A,B]/2>^2. \qquad (3)$$

We will show that the exact lower bound of this uncertainty relation (when eq. (3) becomes equality) can be conveniently used for finding energies of ground and excited states of various quantum systems, as well as for generating simple equations for corresponding wave functions. Some of these results, like an application of the uncertainty principle to a harmonic oscillator or angular momentum, are known and presented below for convenience.

2. **Derivation of the uncertainty relation**

For two arbitrary Hermitian operators $A$ and $B$, state vector $|\psi>$, and constant $\alpha$, the square of the length of the vector $(\alpha A + iB)|\psi>$ should be non-negative:

$$|(\alpha A + iB)|\psi>|^2 = <\psi|(\alpha A - iB)(\alpha A + iB)|\psi> = \alpha^2 <A^2> + \alpha <i[A,B]> + <B^2> \geq 0. \qquad (4)$$

Therefore, the discriminant of the quadratic in $\alpha$ form in eq.(4) should be non-positive:

$$<i[A,B]>^2 - 4<A^2><B^2> \leq 0, \quad \text{or} \quad <A^2><B^2> \geq <i[A,B]/2>^2, \qquad (5)$$

which coincides with eq.(3).

3. **Radial forms of the uncertainty relation**

Instead of the equation (2) for components, it may be convenient using relations for the total momentum and radial coordinate. They can be obtained by inserting in eq.(3) $A = \mathbf{p} = -i\hbar \nabla$ and $B = \mathbf{r} r^{-n}$, where $\mathbf{r} = (x,y,z)$ is the radius-vector and $r$ is its length:

$$<p^2> <r^{2(1-n)}> \geq <[\hbar \nabla, \mathbf{r} r^{-n}]/2>^2. \qquad (6)$$

By calculating the divergence $\nabla \cdot \mathbf{r} r^{-n} = (3-n) r^{-n}$, one obtains

$$<p^2> <r^{2(1-n)}> \geq (\hbar(3-n)/2)^2 <r^{-n}>^2. \qquad (7)$$

For n = 0,1, and 2, one gets useful special forms of this relation:
$$\langle p^2 \rangle \langle r^2 \rangle \geq (3\hbar/2)^2, \tag{7a}$$
$$\langle p^2 \rangle \geq \hbar^2 \langle r^{-1} \rangle^2, \tag{7b}$$
$$\langle p^2 \rangle \geq (\hbar/2)^2 \langle r^{-2} \rangle. \tag{7c}$$

### 4. Harmonic oscillator

The Hamiltonian of the one-dimensional harmonic oscillator is
$$H = (1/2m)\, p_x^2 + (k/2)\, r_x^2. \tag{8}$$
To find the ground state energy we will minimize the total energy $E = \langle H \rangle$ under the constraint imposed by eq.(2). With short notations $x = \langle p_x^2 \rangle$, $y = \langle r_x^2 \rangle$, eq.(2) is $xy \geq (\hbar/2)^2$, $x \geq (\hbar/2)^2(1/y)$, and
$$E = (1/2m)\, x + (k/2)\, y \geq (1/2m)(\hbar/2)^2(1/y) + (k/2)\, y. \tag{9}$$
Minimization with respect to y gives
$$dE/dy = 0 = -(1/2m)(\hbar/2)^2(1/y^2) + (k/2), \quad y = (km)^{-1/2}(\hbar/2). \tag{10}$$
Inserting this value of y and $x = (\hbar/2)^2(1/y)$ into eq.(9), one obtains
$$E_{\min} = (1/2m)(\hbar/2)^2 (km)^{1/2}(\hbar/2)^{-1} + (k/2)(km)^{-1/2}(\hbar/2) = (k/m)^{1/2}(\hbar/2). \tag{11}$$
For three-dimensional symmetric oscillator with the Hamiltonian
$$H = (1/2m)\, p^2 + (k/2)\, r^2 \tag{12}$$
the calculation is exactly the same, except that the constraint (7a) is used instead of eq.(2). The result is
$$E_{\min} = (k/m)^{1/2}(3\hbar/2). \tag{13}$$
The wave function which minimizes the energy also turns to zero the length of the vector in eq.(4):
$$(\alpha A + iB)|\psi\rangle = 0. \tag{14}$$
With zero discriminant in eq.(4) one obtains for 1D oscillator ($A = p_x$, $B = r_x$)
$$\alpha = -\langle i[A,B]/2 \rangle / \langle A^2 \rangle = -(\hbar/2)/\{(\hbar/2)(km)^{1/2}\} = -(km)^{-1/2} \tag{15}$$
and a simple first-order differential equation for the ground-state wave function
$$\{(km)^{-1/2}\hbar\,(d/dr_x) + r_x\}|\psi\rangle = 0 \tag{16}$$
which has the solution
$$|\psi\rangle \propto \exp\{-(r_x^2/2\hbar)(km)^{1/2}\}. \tag{17}$$
One can notice that the calculations we performed for harmonic oscillator closely resemble using creation an annihilation operators for this problem.

### 5. Angular momentum

To find minimum average value of the square of angular momentum $L^2 = L_x^2 + L_y^2 + L_z^2$ at a given value of z-projection, one can use eq.(3) with $A = L_x$, $B = L_y$, and the commutation relation $[L_x, L_y] = i\hbar L_z$:
$$\langle L_x^2 \rangle \langle L_y^2 \rangle \geq \langle i[L_x, L_y]/2 \rangle^2 = (\hbar/2)^2 \langle L_z \rangle^2. \tag{18}$$
For the states with definite values of $\langle L_z \rangle = \hbar m$ this gives $\langle L_x^2 \rangle = \langle L_y^2 \rangle \geq (1/2)\hbar^2|m|$ and
$$\langle L^2 \rangle = \langle L_x^2 \rangle + \langle L_y^2 \rangle + \langle L_z^2 \rangle \geq \hbar^2|m| + (\hbar m)^2 = \hbar^2 l(l+1) = \langle L^2 \rangle_{\min}, \tag{19}$$
where $l = |m|$.

### 6. Hydrogen-like atoms

It will be convenient using atomic units: $\hbar = m = e' = 1$ and $e'^2 = e^2/(4\pi\varepsilon_0)$ in SI. With these units, the energy is measured in *hartree* and the length in *bohrs* (1 *hartree* = 27.211 eV, 1 *bohr* = 0.52918 Å). The Hamiltonian of a hydrogen-like atom is

$$H = (1/2)p^2 - Zr^{-1}, \tag{20}$$

where $Z$ is the nucleus charge. By using $x = \langle p^2 \rangle$, $y = \langle r^{-1} \rangle$ and eq.(7b), which takes the form $x \geq y^2$, one obtains

$$E = \langle H \rangle = (1/2)x - Zy \geq (1/2)y^2 - Zy. \tag{21}$$

Minimization gives

$$dE/dy = 0 = y - Z, \quad y = Z, \tag{22}$$

and the minimum energy

$$E_{min} = (1/2)Z^2 - Z^2 = -(1/2)Z^2. \tag{23}$$

The equation for the ground-state wave function can be obtained in the same way it was done in eqs.(14-16) by using $A = \mathbf{p}$ and $B = \mathbf{r}r^{-1}$:

$$\alpha = -\langle i[A,B]/2 \rangle / \langle A^2 \rangle = -\langle i[\mathbf{p},\mathbf{r}r^{-1}]/2 \rangle / \langle p^2 \rangle = -y/y^2 = -Z^{-1}, \tag{24}$$

$$0 = (\alpha A + iB)|\psi\rangle = (-Z^{-1}\mathbf{p} + i\mathbf{r}r^{-1})|\psi\rangle, \tag{25}$$

which leads to the first-order differential equation

$$(Z^{-1}\nabla + \mathbf{r}r^{-1})|\psi\rangle = 0 \tag{26}$$

with the solution

$$|\psi\rangle \propto \exp(-Zr). \tag{27}$$

To find the energies of states with non-zero angular momentum, one can use eq.(3) with $A = \boldsymbol{\sigma} \cdot \mathbf{p}$ and $B = \boldsymbol{\sigma} \cdot \mathbf{r}r^{-1}$, where the vector $\boldsymbol{\sigma} = (\sigma_x, \sigma_y, \sigma_z)$ is the Pauli spin operator. Eq.(3) then becomes

$$\langle p^2 \rangle \geq \langle i[\boldsymbol{\sigma} \cdot \mathbf{p}, \boldsymbol{\sigma} \cdot \mathbf{r}r^{-1}]/2 \rangle^2 = \langle r^{-1}(1 + \boldsymbol{\sigma} \cdot \mathbf{L}) \rangle^2 = \langle r^{-1}(1 + l) \rangle^2, \tag{28}$$

where we used the algebra of Pauli matrices ($\sigma_x^2 = 1$, $[\sigma_x, \sigma_y] = 2i\sigma_z$, $\sigma_x \sigma_y + \sigma_y \sigma_x = 0$, etc.). Starting again with eq.(21), one obtains

$$E = \langle H \rangle = (1/2)x - Zy \geq (1/2)y^2(1 + l)^2 - Zy. \tag{29}$$

Minimization gives

$$dE/dy = 0 = y(1 + l)^2 - Z, \quad y = Z(1 + l)^{-2}, \tag{30}$$

and the minimum energy

$$E_{min} = (1/2)Z^2(1 + l)^{-2} - Z^2(1 + l)^{-2} = -(1/2)Z^2(1 + l)^{-2} = -Z^2/2n^2, \tag{31}$$

where $n = 1 + l$.

## 7. He

The Hamiltonian for the Helium atom is

$$H = (1/2)p_1^2 + (1/2)p_2^2 - 2r_1^{-1} - 2r_2^{-1} + r_{12}^{-1}. \tag{32}$$

By introducing $x = \langle p_1^2 \rangle = \langle p_2^2 \rangle$, $y = \langle r_1^{-1} \rangle = \langle r_2^{-1} \rangle$, and $z = \langle r_{12}^{-1} \rangle$, one can write the average energy as

$$E = x - 4y + z. \tag{33}$$

The constraint given by eq.(7b) is $x \geq y^2$. $z$ and $y$ are not independent, so minimization cannot be done by independently varying them. The simplest relation between $z$ and $y$, consistent with the virial theorem [6], is $z = Cy$, where $C$ is a constant. With this linear relation,

$$E = x - y(4-C) \geq y^2 - y(4-C). \tag{34}$$

Minimization gives $y = (4-C)/2$ and

$$E_{min} = -(4-C)^2/4. \tag{35}$$

If one assumes a special functional form of the wave function, $C$ can be calculated explicitly. As an example, for a hydrogen-like functions

$$|\psi\rangle \propto \exp\{-\beta(r_1 + r_2)\}, \tag{36}$$

$C = 5/8$ and, from the eq.(35), $E_{min} = -(27/16)^2 = -2.848$ or $-77.50$ eV. (The best variational solution [7], truncated to four digits, is $-2.904$). Expectedly, one would obtain exactly the same result by using a conventional variation method with the separable hydrogen-like functions in eq.(36) [8].

## 8. Conclusion

When the Hamiltonian is a sum of several non-commuting operators, its expectation value (average energy) can be minimized by varying the average values of these operators under the constraints imposed by the uncertainty principle. Such procedure may suggest convenient methods of finding the ground state energy, and it can also be useful for finding the energies of other stationary states and wave functions.

[1] W. Heisenberg, *Über den anschaulichen Inhalt der quantentheoretischen Kinematik und Mechanik*, Zeitschrift für Physik **43**, 172–198 (1927). doi:10.1007/BF01397280. English translation: J. A. Wheeler and H. Zurek, *Quantum Theory and Measurement*, Princeton Univ. Press, pp. 62–84 (1983).
[2] E. H. Kennard, *Zur Quantenmechanik einfacher Bewegungstypen*, Zeitschrift für Physik **44,** 326 (1927). doi:10.1007/BF01391200.
[3] H. Weyl, *Gruppentheorie Und Quantenmechanik*. Leipzig: Hirzel (1928).
[4] H. P. Robertson, *The Uncertainty Principle*, Phys. Rev. **43**, 163–64 (1929).
[5] E. Schrödinger, *Zum Heisenbergschen Unschärfeprinzip*, Sitzungsberichte der Preussischen Akademie der Wissenschaften, Physikalisch-mathematische Klasse **14**, 296–303 (1930).
[6] P.-O. Löwdin, *Scaling problem, virial theorem, and connected relations in quantum mechanics*, J. Mol. Spectr. **3**, 46–66 (1959). doi:10.1016/0022-2852(59)90006-2.
[7] G. W. F. Drake, Z.-C. Yan, *Energies and relativistic corrections for the Rydberg states of helium: Variational results and asymptotic analysis*, Phys. Rev. A **46**, 2378 (1992).
[8] I. N. Levine, *Quantum Chemistry* (Prentice Hall, New Jersey, 2000).